\documentclass[12pt]{article}
\usepackage{multicol}

\usepackage[labelfont=bf,textfont=footnotesize]{caption}
\DeclareCaptionType{fig}[FIGURE][List of mytype]

\def\beq{\begin{equation}}
\def\eeq#1{\label{#1}\end{equation}}
\def\eeqn{\end{equation}}

\def\beqa{\begin{eqnarray}}
\def\eeqa#1{\label{#1}\end{eqnarray}}
\def\eeqan{\end{eqnarray}}

\let\bar=\overbar

\def\Dslash{\not{\hbox{\kern-4pt $D$}}}
\def\dslash{\not{\hbox{\kern-2pt $\del$}}}

\def\msb{{\bar{\ssstyle M \kern -1pt S}}}

\usepackage{fancyhdr,graphicx}
\def\Title#1{\begin{center} {\Large {\bf #1} } \end{center}}

\begin{document}

\Title{Phase diagram of strongly interacting matter under strong
magnetic fields}

\bigskip\bigskip


\begin{raggedright}

Pablo G. Allen$^1$ and Norberto N. Scoccola$^{1,2,3}$\\
{\it $^1$Department of Theoretical Physics, GIyA,  CNEA,
Libertador 8250, 1429.\\
$^2$CONICET, Av. Rivadavia 1917, 1033.\\
$^3$Universidad Favaloro, Sol\'is 453, 1078, Buenos Aires,
Argentina.}
\bigskip\bigskip

\end{raggedright}

\section{Introduction}

The understanding of the behaviour of strongly interacting matter
at finite temperature and density is of fundamental interest and
has applications in cosmology, in the astrophysics of neutron
stars and in the physics of relativistic heavy ion collisions.
Given the possible existence of strong magnetic fields in the
mentioned situations, their effect on the QCD phase diagram has
recently become a topic of increasing interest\cite{book}. Here,
we report on the study of this issue in the framework of the two
flavor Nambu-Jona-Lasinio model with Polyakov loop (PNJL)
\cite{PNJL} and an extension of it, the so-called entangled PNJL
model (EPNJL)\cite{EPNJL}. Previous analyses at finite temperature
can be found in Ref.\cite{Fukushima:2010fe}.

\section{Formalism}

Our starting point for the study of quark matter is the PNJL
model, which is constructed by incorporating the Polyakov Loop
(PL) into the finite temperature and chemical potential
Nambu-Jona-Lasinio (NJL) model\cite{reports}. Since the
model under consideration is not renormalizable, we need to
specify a regularization scheme. Here, we introduce a sharp
cut-off in 3-momentum space, only for the divergent ultra-violet
integrals. The Euclidean PNJL action coupled to the EM field reads
\begin{equation}
S_E = \int d^4 x \left\{
\overline{\psi} (- i \gamma^{\mu} D_\mu  + m_0) \psi -
G\left[(\overline{\psi}\psi)^2  +  (\overline{\psi} i\tau\gamma_{5}\psi)^2 \right]
+ {\cal U}(\Phi[{\cal G}(x)])\right\},
\label{ionflux}
\end{equation}
where $m_0$ is the current mass and G is a coupling constant.
Together with the cut-off $\Lambda$, they completely determine the
model. Two sets of parameters were used. For Set 1: $m_0=5.6$ MeV,
$\Lambda=587.9$ MeV, $G\Lambda^{2}=2.44$, and for Set 2: $m_0=5.5$
MeV, $\Lambda=631.5$ MeV, $G\Lambda^{2}=2.19$. The coupling of the
quarks to the (electro)magnetic field ${\cal A}_\mu$ and the gluon
field ${\cal G}_\mu$ is implemented  {\it via} the covariant
derivative $D_{\mu}=\partial_\mu - i q_f {\cal A}_{\mu}-i {\cal
G}_\mu$ where $q_f$ represents the quark electric charge ($q_u/2 = -q_d = e/3$).
We consider a  static and constant magnetic field
in the $z$ direction, ${\cal A}_\mu=\delta_{\mu 2} x_1 B$.
Concerning the gluon fields, we assume that quarks move on a
constant background field ${\cal G}_\mu= \delta_{\mu4} \phi$. Then the traced
Polyakov loop, which in the infinite quark mass limit can be taken
as an order parameter of confinement, is given by
$\Phi=\frac{1}{3} {\rm Tr}\, \exp( i \phi/T)$. We work in the
so-called Polyakov gauge, in which $\phi = \phi_3 \lambda_3 + \phi_8
\lambda_8$. This leaves only two independent variables, $\phi_3$
and $\phi_8$. In the case of $\mu=0$ the
traced Polyakov loop in the Mean Field Approximation (MFA) is
expected to be a real quantity implying $\phi_8=0$, a condition
that we assume to be valid also for finite real $\mu$. The MFA traced Polyakov
loop reads then $ \Phi = \Phi^* = \left[ 1 + 2 \,\cos
\left(\phi_3/T\right)\right]/3$. To proceed we need to specify the
explicit form of the Polyakov loop effective potential
${\cal{U}}(\Phi ,T)$. Here we consider~\cite{Roessner:2006}
\begin{eqnarray}
{\cal U}(\Phi,T)= - \frac{1}{2} a(T)  \Phi \Phi^* + b(T) \ln [1 -
6\Phi\Phi^* + 4 (\Phi^3 + \Phi^{*3}) - 3 (\Phi\Phi^*)^2 ],
\end{eqnarray}
where $a(T) = a_0 +a_1 \left(T_0/T\right) + a_2\left(T_0/T\right)^2$ and
$b(T) = b_3\left(T_0/T\right)^3$.
The values of the constants $a_i, b_3$ can be fitted to pure gauge
lattice QCD (LQCD) results, leading to $ a_0 = 3.51\ ,a_1 = -2.47\ , a_2
= 15.2\ , b_3 = -1.75$~\cite{Roessner:2006}. The scale
parameter $T_0$ corresponds in principle to the
deconfinement transition temperature in the pure Yang-Mills
theory, $T_0 = 270$~MeV. However, it has been argued that in the
presence of light dynamical quarks this temperature scale should
be adequately reduced~\cite{Schaefer:2007pw}. Thus, we
also consider values $T_0 \sim 200$ MeV in our calculations.

In the standard PNJL model the quark-quark coupling constant $G$ is
independent of the PL. To account for further
correlations between the quark and colour sector, a PL dependent
$G(\Phi)$ can be introduced, leading to the EPNJL model. Namely,
\begin{equation}
G(\Phi)=  \left[ 1 - \alpha_{1}\Phi\Phi^* - \alpha _{2}( \Phi^3 +
\Phi^{*3}) \right] G
\end{equation}
The choice $\alpha_1= \alpha_2=0.2$
reproduces the LQCD phase diagram at
imaginary $\mu$\cite{EPNJL}.

Finally, using the Matsubara formalism to account for finite $T$
and $\mu$ in the quark sector, the MFA thermodynamical potential
for the models under consideration read
\begin{eqnarray}
\Omega_{MFA}(M,\Phi)&=& \frac{(M-m_0)^2}{4G(\Phi)} + {\cal U}(\Phi,T)-
\frac{N_cN_f}{\pi^2} \int_{0}^{\Lambda}dp \ p^2 \sqrt{p^2+M^2}
\nonumber\\
&-& \!\!\! \frac{N_c}{2\pi^2} \! \sum_{f=u,d}(q_f B)^2
\left[ \zeta'(-1,x_f) + \frac{x_f^2}{4} - \frac{1}{2} (x_f^2 -x_f)\log x_f \right]
\nonumber\\
&-& \!\!\!  \frac{T}{2\pi}  \! \sum_{s,k,c,f}  \! \alpha_k |q_f| B  \!
\int \! \frac{dp_z}{2\pi}
\ln  \!
\left[ 1 + \exp\left( -\frac{ E_f(p_z,k) +s\ \mu + i\phi_c }{T}\right) \right]\! ,
\end{eqnarray}
where $s=\pm 1$, $\alpha_k = 2- \delta_{k0}$,
$E_f(p_z,k)=\sqrt{M^2 + p_z^2 + 2 k |q_f| B}$, and
$\phi_c=\phi_3,\phi_3, -2 \phi_3$ for $c=r,g,b$, respectively. In
addition, $x_f= M^2/(2 |q_f| B)$. The dressed quark mass $M$ and
the PL are found as solution of the gap equations,
$\partial\Omega_{MFA}/\partial M = \partial \Omega_{MFA}/\partial
\Phi =0$. Once the solutions are found, chiral and deconfinement
critical temperatures for crossover transitions are defined as the
peaks of the corresponding susceptibilities $\chi_{ch}=
\partial M/\partial T$ and  $\chi_\Phi =
\partial\Phi/\partial T$.

\section{Results}

We consider first the behavior of the dressed quark mass as a
function of the magnetic field at $T=\mu=0$. As shown in Fig.1(a)
the presence of a magnetic field strengthens the breaking of the
symmetry, or, equivalently, stabilizes the chiral condensate,
phenomena known as magnetic catalysis. If we now include the
effect of finite temperature, chirally restored and deconfined
phases are both found to exist for sufficiently high temperatures.
The behavior of the corresponding critical temperatures at $\mu=0$
as functions of $B$ is displayed in Figs.1(b,c). Without
entanglement, the deconfinement temperature depends only weakly on
the magnetic field and the splitting between both transitions
increases with $B$. However, in the entangled case, both
transitions increase together. For a higher value of $T_0$, the
entanglement becomes less effective for high magnetic fields. In
any case, we find that the present models predict an increase of
the critical temperatures as $B$ increases, a result which seems
at variance with the most recent LQCD results\cite{baliJHEP2012}.
\begin{figure} [h]
\hspace*{.5cm}
\includegraphics[width=0.7 \textwidth]{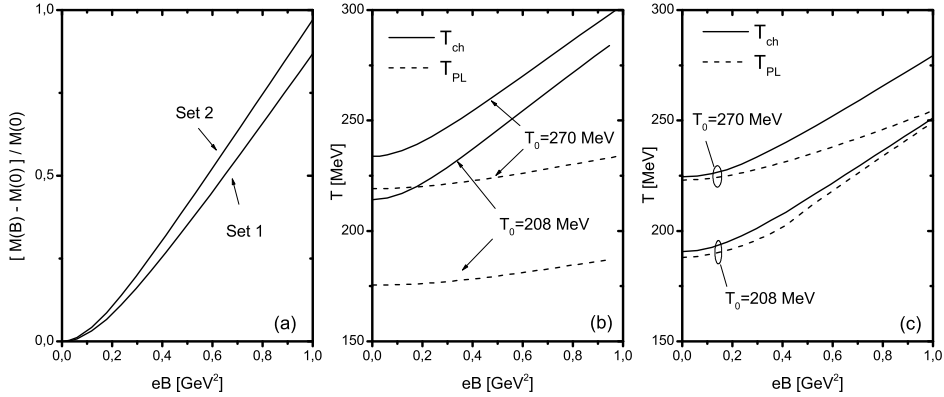}
 \caption{(a)Magnetic Catalysis: M increases with B, for both Sets.
 (b) Critical Temperatures PNJL, Set 1. (c) Critical Temperatures E-PNJL, Set 1} 
 \end{figure}
In Fig.2, the effect of the magnetic field on the phase diagram is
shown for Set~1. The magnetic field shifts the crossover chiral
transition upwards in all cases. In the EPNJL model, due to
additional correlations between the quark and gluon sectors, this
affects the deconfinement transition. On the other hand, in the
PNJL model such transition is practically independent of $B$. The
behaviour of critical $\mu$ at $T=0$ is non trivial, first
diminishing with $B$ and then increasing. A minimum value is
attained near $eB=0.3$ GeV$^2$.
\begin{figure} [h]
 \begin{center}
\includegraphics[width=0.6 \textwidth]{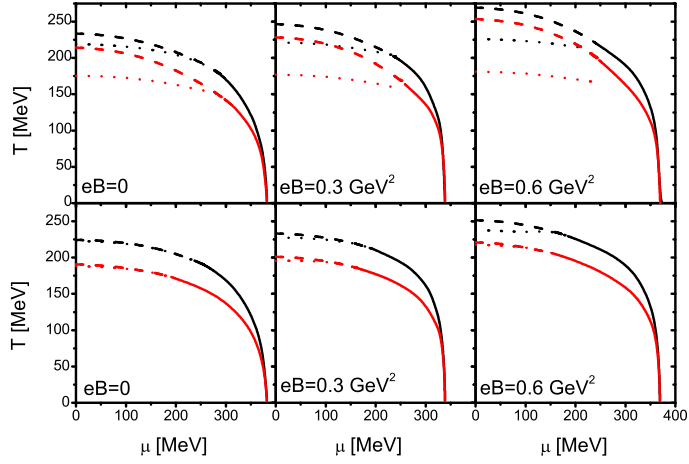}
\caption{Set 1: Phase Diagrams for several values of the magnetic field.
 Upper panels correspond to the PNJL model, while lower ones to the EPNJL model.
 Dashed lines correspond to the chiral restoration crossover while full line to the 1st order
 one. Dotted lines correspond to the deconfinement transition. In each panel, upper (black) lines
 correspond to $T_0=270$ MeV while lower (red) ones to $T_0=208$ MeV.}
 \end{center}
 \end{figure}
The behaviour of the CEP is seen in Fig.3. $T_{CEP}$ increases with magnetic field,
while $\mu_{CEP}$ tends to decrease, presenting small oscillations. Similar results
have been found in the NJL model\cite{Avancini:2012ee}.
\begin{figure} [h]
 \begin{center}
\includegraphics[width=0.8 \textwidth]{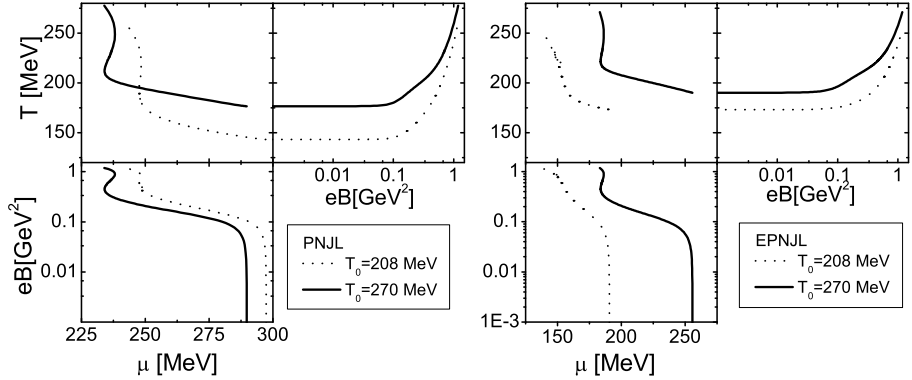} \caption{Set 1: Position of CEP as a function of Magnetic Field } 
 \end{center}
 \end{figure}

In the phase diagram of Set~2 (Fig.4), it is seen that at values
around $eB=0.1$ GeV$^2$ chiral restoration at low $T$ occurs in
several steps: a main transition and two secondary ones, that also
turn to crossovers at a CEP. In the EPNJL case (lower panel),
there is more than one CEP in the main transition. In Fig.5, the
response to magnetic field of all the encountered CEPs is shown.
In the PNJL case (left panel), it is seen that there is one CEP at
$B=0$, that moves towards the $T=0$ axis when $B$ increases, while
the CEP corresponding to one of the secondary transitions moves to
higher $T$ values and remains for higher $B$, turning into the
main transition CEP. The situation in the EPNJL model(right panel)
is much more complicated: at $eB=0.079$ GeV$^2$ a pair of CEPs is
formed, one of them disappearing after encountering a CEP of one
of the secondary transitions, while the other one disappears in
the $T=0$ axis. The CEP at $B=0$ remains as the main CEP for all
values of $B$.

 \begin{figure} [h]
 \begin{center}
 \includegraphics[width=0.6 \textwidth]{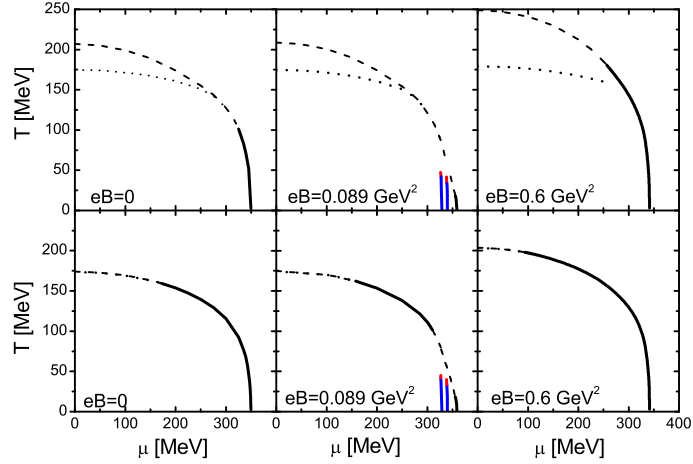}
 \caption{Set 2: Phase Diagrams for several values of the magnetic field.
 Upper panels correspond to the PNJL model, while lower ones to the EPNJL model.
 Dashed lines correspond to the chiral restoration crossover while full line to the 1st order
 one. Dotted lines correspond to the deconfinement transition. Results using
 $T_0=212 (190)$ MeV for PNJL (EPNJL) are shown.}
 \end{center}
 \end{figure}

 \begin{figure} [h]
 \begin{center}
 \includegraphics[width=0.8 \textwidth]{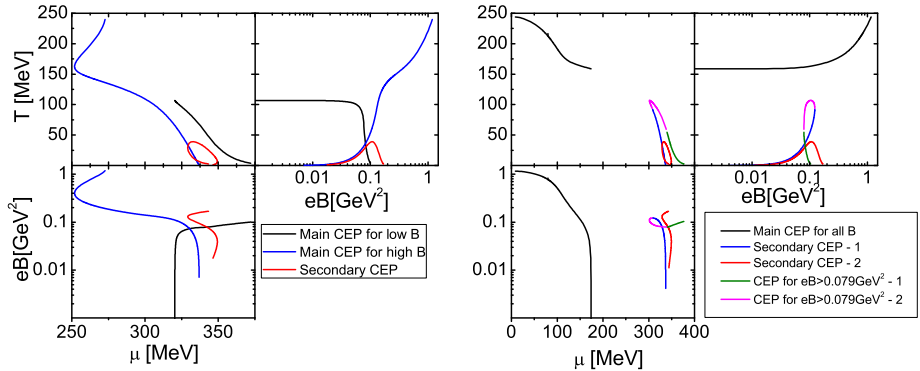}
 \caption{Set 2: Position of CEP as a function of magnetic field. Values
 of $T_0$ as in Fig.4.} 
 \end{center}
 \end{figure}

\section{Conclusions}
We have analyzed the effect of an intense magnetic field on the
phase diagram of strongly interacting matter as described by
(E)PNJL-type models. These models provide a simultaneous dynamical
description of the deconfinement and chiral transitions. They are
able to describe the enhancement of the chiral condensate with $B$
at $T=\mu=0$. However, as most of the present available models, in
their present version they fail to reproduce the inverse magnetic
catalysis at finite temperature recently found in lattice QCD. In
the EPNJL model there is no splitting at $\mu=0$ between chiral
restoration and deconfinement transitions as functions of $B$.
Similarly for a given $B$ both transitions coincide up to the
critical point. The detailed form of the phase diagram depends,
particularly at low $T$, on the quark sector parametrization. For
parametrizations leading to a $T=\mu=0$ dressed quark mass smaller
than $\simeq 350 MeV$ (as in Set 2) there is a quite rich
structure due to the subsequent population of the Landau levels as
$\mu$ increases. In particular, several CEPs are found.

\bigskip
{\it This work has been partially funded by CONICET (Argentina) under grants \#
PIP 00682, and by ANPCyT (Argentina) under grant \# PICT11 03-113.}

\end{document}